\documentclass[]{spie}  %>>> use for US letter paper
\usepackage{amsmath,amsfonts,amssymb}
\usepackage{graphicx}
\usepackage[colorlinks=true, allcolors=blue]{hyperref}

\title{Laboratory demonstration of the triple-grating vector vortex coronagraph}

\author[a,b]{David S. Doelman}
\author[a]{Mireille Ouellet}
\author[c]{Axel Potier}
\author[c]{Garreth Ruane}
\author[d]{Kyle van Gorkom}
\author[d]{Sebastiaan Y. Haffert}
\author[d]{Ewan S. Douglas}
\author[a]{Frans Snik}

\affil[a]{Leiden Observatory, Leiden University, P.O. Box 9513, 2300 RA Leiden, The Netherlands}
\affil[b]{SRON Netherlands Institute for Space Research, Niels Bohrweg 4, 2333 CA, Leiden, The Netherlands}
\affil[c]{Jet Propulsion Laboratory, California Institute of Technology, Pasadena, California 91109, USA}
\affil[d]{University of Arizona, Steward Observatory, Tucson, Arizona, United States}

\authorinfo{Further author information: Send correspondence to D.S.D\\D.S.D.: E-mail: D.S.Doelman@SRON.nl}

\begin{document} 
\maketitle

\begin{abstract}
The future Habitable Worlds Observatory aims to characterize the atmospheres of rocky exoplanets around solar-type stars. The vector vortex coronagraph (VVC) is a main candidate to reach the required contrast of $10^{-10}$. However, the VVC requires polarization filtering and every observing band requires a different VVC. The triple-grating vector vortex coronagraph (tgVVC) aims to mitigate these limitations by combining multiple gratings that minimize the polarization leakage over a large spectral bandwidth. In this paper, we present laboratory results of a tgVVC prototype using the In-Air Coronagraphic Testbed (IACT) facility at NASA's Jet Propulsion Laboratory and the Space Coronagraph Optical Bench (SCoOB) at the University of Arizona Space Astrophysics Lab (UASAL). We study the coronagraphic performance with polarization filtering at 633 nm and reach a similar average contrast of $2\times10^{-8}$ between 3-18 $\lambda/D$ at the IACT, and $6\times10^{-8}$ between 3-14 $\lambda/D$ at SCoOB. We explore the limitations of the tgVVC by comparing the testbed results. We report on other manufacturing errors and ways to mitigate their impact.
\end{abstract}


\section{INTRODUCTION}
\label{sec:intro}  % \label{} allows reference to this section

Detecting and characterizing the atmospheres of rocky exoplanets in the habitable zone of solar-type stars is recognized as the main scientific goal of NASA's future flagship mission, the Habitable Worlds Observatory. 
To achieve this immense goal, it is necessary to achieve $10^{-10}$ contrast over 20\% bandwidth at separations close to the inner working angle of a 6.5 m telescope \cite{national2021pathways}. 
It has been recognized that most components of high-contrast imaging systems are close but do not yet have the technological maturity to meet these requirements \cite{clery2023future}\footnote{\url{https://science.nasa.gov/astrophysics/programs/gomap}}.
At the heart of these systems is the coronagraph, a combination of optical masks that reject on-axis starlight in a region called the dark zone by many orders of magnitude while light from off-axis sources is only slightly affected.
So far, only Lyot coronagraphs have demonstrated $4\times10^{-10}$ in an annular dark zone at 10\% bandwidth \cite{seo2019testbed}, and the same contrast for 20\% bandwidth in a D-shaped dark zone with larger inner working angle \cite{allan2023demonstration}. 
While this is close, a significant push is still necessary to reach the required performance to directly image Earth-twins in the habitable zone. \\
The vortex coronagraph \cite{foo2005optical} is another promising coronagraphic architecture under study for the Habitable Worlds Observatory. 
This coronagraphic architecture theoretically has a much improved performance for unobstructed apertures.
The vortex coronagraph consists of a mask with an azimuthal phase ramp in the focal plane that diffracts the on-axis star light outside of the pupil in the next pupil plane, where it is blocked by a Lyot stop.
Two ways to manufacture these masks in the focal plane are generally considered. 
First, there is the mask based on classical phase, where the phase is accrued by changing the optical path length, making it chromatic. 
Manufacturing of such a mask can be done by etching variations in height \cite{lee2006experimental} or by optimizing pillar shapes and densities in meta-materials \cite{Konig2023metasurface}.
Because the phase is chromatic, multiple solutions are considered to improve the chromatic behaviour.
Examples are the changing of the azimuthal phase of the mask \cite{ruane2019scalar,galicher2020family,desai2023achromatizing,desai2023laboratory}, or by selecting pilar heights in meta-materials such that the phase ramp is the same for a wider wavelength range, except with a overall phase offset \cite{palatnick2023imaging,Konig2023metasurface}.
Overall, these methods that use classical phase have decent performance but chromaticity continues to be an issue.\\
A second type of mask is based on the achromatic geometric phase. 
Geometric phase is accrued when circularly polarized light goes through a half-wave retarder with a rotated fast-axis \cite{pancharatnam1956generalized,Berry1987}.
The geometric phase only depends on the fast-axis orientation and the handedness of the circular polarization coming in and is independent of wavelength.
The vector vortex coronagraph (VVC) based on the geometric phase is therefore a flat patterned half-wave retarder where the fast-axis rotates in the azimuthal direction \cite{mawet2005annular, mawet2009optical}. 
When the retardance is not half-wave, a fraction of the light does not acquire any phase and is unaffected by the mask, i.e. the polarization leakage.
The retardance therefore determines the diffraction efficiency of the mask. 
Polarization leakage of the VVC leads to an attenuated copy of the star on the focal plane, resulting in a performance loss that can only be solved by filtering one circular polarization state. \\
In this paper, we describe how combining multiple grating patterns with a VVC can reduce the impact of polarization leakage by many orders of magnitude in Section \ref{sec:mg}.
Next, we describe the design and manufacturing of two prototype triple-grating VVCs in Section \ref{sec:proto} and characterize their performance on the In-Air Coronagraphic Testbed at NASA/JPL and the Space Coronagraph Optical Bench at the University of Arizona in Section \ref{sec:contrast}. We discuss some manufacturing limitations in Section \ref{sec:discussion} and present our conclusions in Section \ref{sec:conclusion} 

\section{Minimizing polarization leakage by with multi-grating stacks}
\label{sec:mg}
To overcome limitation of polarization leakage and the need for leakage filtering when using the VVC, we introduced the multi-grating concept in Doelman et al. 2020 \cite{doelman2020minimizing}, which has already successfully been tested on-sky with the vector Apodizing Phase plate coronagraph \cite{doelman2020minimizing,doelman2021vector}.
\begin{figure} [ht]
   \begin{center}
     \includegraphics[width = \linewidth]{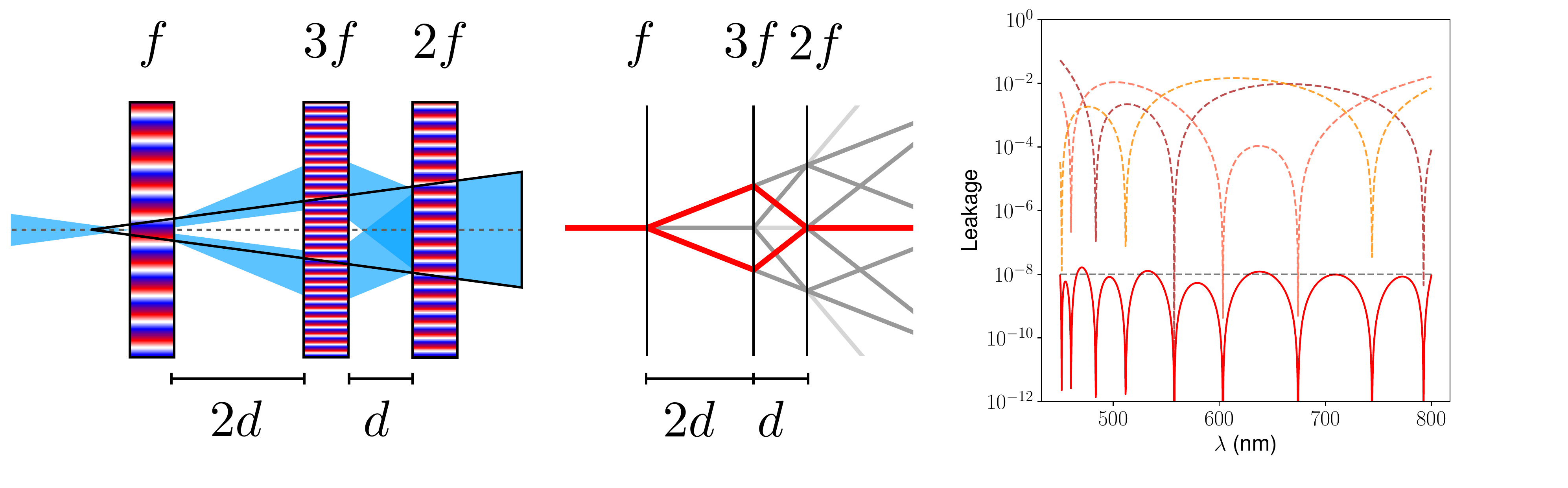}
   \end{center}
   \caption[example] 
%>>>> use \label inside caption to get Fig. number with \ref{}
   { \label{fig:f1} Concept of the triple-grating vector vortex coronagraph. \textit{Left:} The incoming unpolarized light is diffracted in two main directions with opposite circular polarization, which occur equally in unpolarized light. The beams are diffracted in the opposite direction by the next grating and back on-axis by the third. This is only enabled by the polarization gratings diffracting circularly polarized light with extreme efficiency ($>98\%$) into the $\pm 1$ orders. \textit{Middle:}  We show the main beams in red, while all leakage beams are shown in gray. The gray scale indicates the relative intensity of the leakage terms. \textit{Right:} The simulated on-axis polarized leakage for the tgVVC created with liquid-crystal layers with three sublayers. This idealized simulation shows an on-axis leakage lower than $10^{-8}$ over the full visible wavelength-range, resulting in a $10^{-10}$ intensity at the first Airy ring. The dotted lines indicate the polarization leakage for each individual grating. Adopted from Doelman et al. 2020 \cite{doelman2020minimizing}}
   \end{figure} 

    \begin{figure} [ht]
   \begin{center}
     \includegraphics[width = \linewidth]{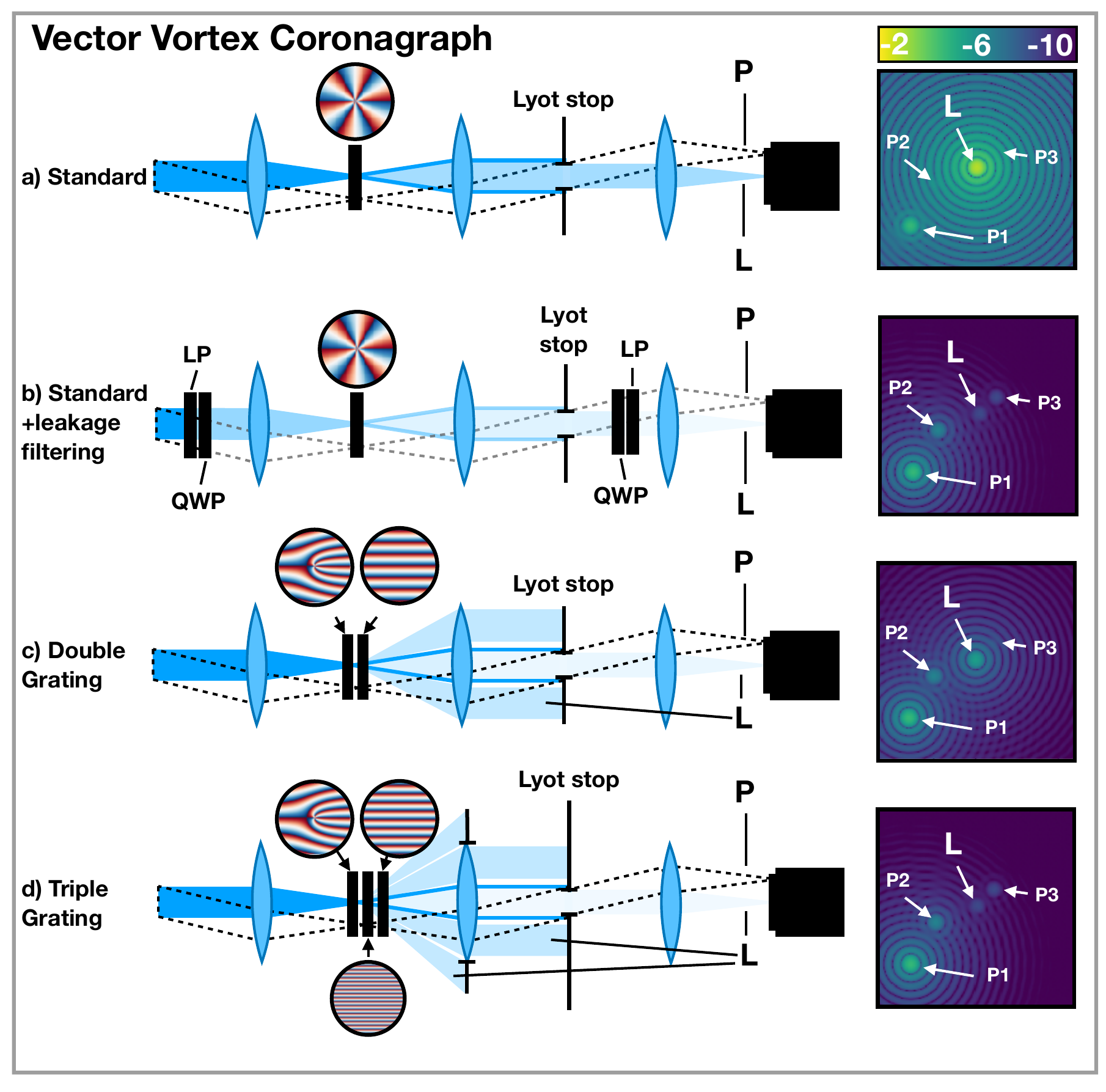}
   \end{center}
   \caption[example] 
%>>>> use \label inside caption to get Fig. number with \ref{}
   { \label{fig:f2} Conceptual comparison of a non-perfect VVC with polarization leakage (L) with and without filtering to two multi-element VVCs. The focal-plane images are on logarithmic scale. We show three planets, where P1 has a contrast of $1\times10^{-5}$, P2:
$1\times10^{-7}$, and P3: $1\times10^{-9}$. The two closest planets (P2,3) are mostly hidden by the polarization leakage for the VVC. The polarization leakage is diffracted by the double- or triple- grating patterns, resulting in reduced polarization leakage on-axis and the appearance of P2 and P3. Adapted from Doelman et al. 2020. \cite{doelman2020minimizing}}
   \end{figure} 
The core idea is that by combining phase ramps with the VVC pattern, the polarization leakage can be diffracted off-axis and no polarization filtering is required.
   We make use of phase ramps from polarization gratings (PGs), hence the name multi-grating. 
PGs are another type of geometric phase mask and consists of a half-wave retarder with continuously rotating fast-axis in one direction \cite{Oh2009}.
The rotating fast-axis creates the phase ramp in geometric phase, and thus optimally separates left- and right-handed circularly polarized light in the $\pm 1$ diffraction order.
Unpolarized stellar light is diffracted equally in both orders, while deviations from half-wave lead to a zero-order term from the polarization leakage.
The concept of combining three gratings to make a triple-grating VVC is shown in Figure \ref{fig:f1}. 
By combining three polarization gratings with different grating frequencies, f, in a 1:3:2 ratio, we minimize the on-axis polarization leakage by diffracting most leakage off-axis.
 The main beams are recombined after they are separated by the first grating, and maintain a high throughput. 
Triple-grating VVCs can be manufactured using liquid-crystal technology.
Direct-write\cite{Miskiewicz2014} is the only method to accurately pattern the VVC and PG patterns, while using self-aligning multi-layer stacks of liquid-crystals to achromatize the retardance\cite{Komanduri2013} results in polarization leakages $<1\%$ over a spectral range of a few hundered nanometer. 
In the right panel of Figure \ref{fig:f1} we demonstrate that combining three gratings, each manufactured from three-layered liquid-crystal films, can suppress the on-axis leakage by 8 orders of magnitude.\\
The implementation of the multi-grating VVC (mgVVC) on system level and the comparison with a normal vortex is shown in Figure \ref{fig:f2}. 
A normal VVC requires polarization filtering to reach space-based contrasts levels better than $10^{-6}$ as the on-axis leakage PSF is dominant.
For a double-grating VVC (dgVVC) with three-layered liquid-crystal films, the polarization leakage can be suppressed more than six orders of magnitude over a bandwidth of 20\%.
The diffracted polarization leakage is blocked by the Lyot stop. 
Half of the diffracted and blocked polarization leakage has been diffracted by the first grating and leaked through the second grating, and will form a ring of fire outside the Lyot stop. 
To minimize the ring of fire tail leaking through the Lyot stop opening, the grating periods must be minimized.
For the triple-grating VVC (tgVVC), more diffraction orders are generated and these have increased angles. 
Some of these will not be imaged by the lens for slow optical systems, and should be blocked.
However, the on-axis leakage is also reduced by at least two additional orders of magnitude, as was demonstrated in Figure \ref{fig:f1}.
Ultimately, these idealized simulated performances need to be tested and we will describe the manufacturing of the prototype next.

\section{The triple-grating vortex prototype}
\label{sec:proto}
The first two triple-grating vortex prototypes have been manufactured by ImagineOptix in the second half of 2021.
Their design is based on requirements and the F-number of the NASA/JPL In-Air Coronagraphic Testbed (IACT) and manufacturing limitations. 
 \begin{figure} [ht]
   \begin{center}
     \includegraphics[width = \linewidth]{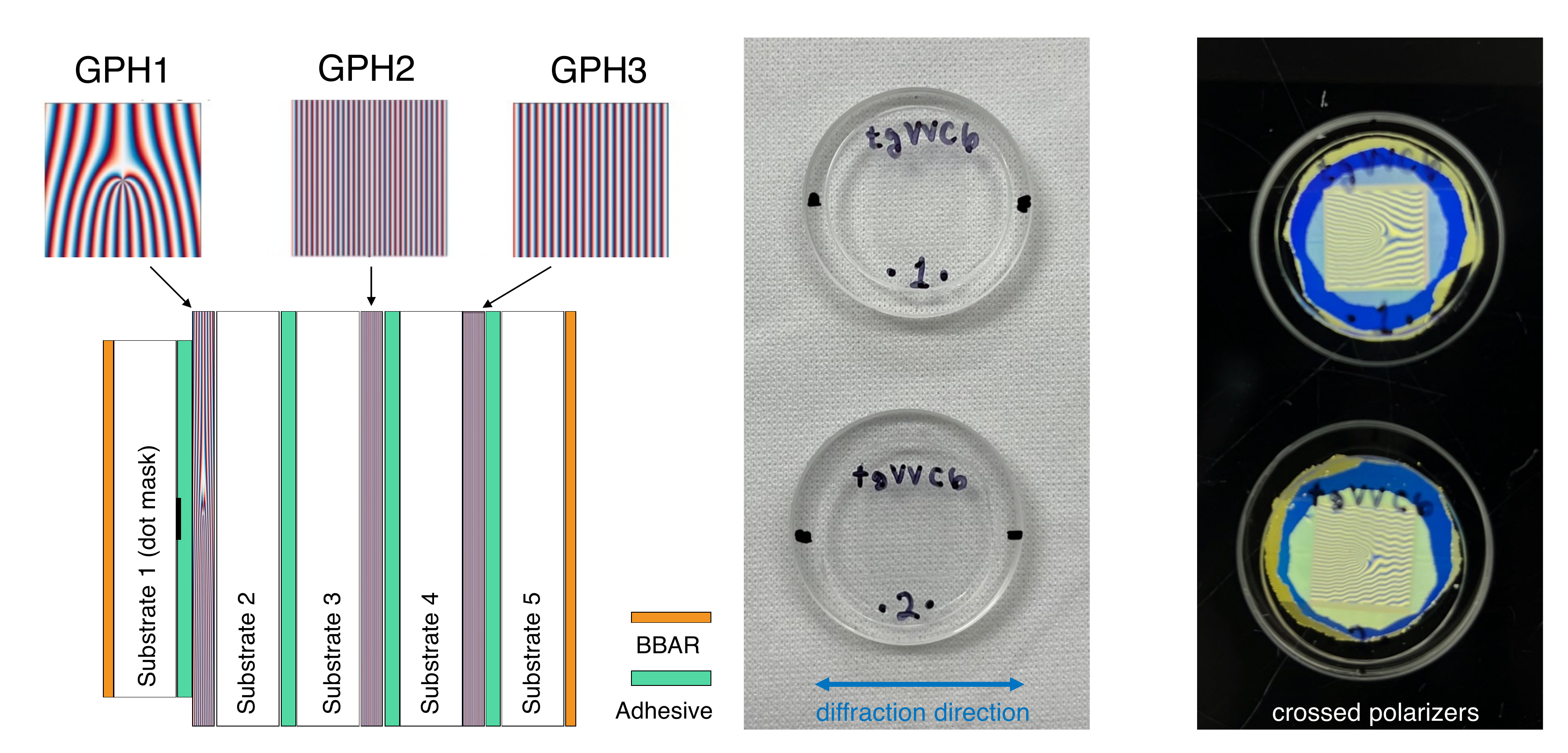}
   \end{center}
   \caption[example] 
%>>>> use \label inside caption to get Fig. number with \ref{}
   { \label{fig:f3} Schematic and images of the tgVVC prototypes manufactured by ImagineOptix in 2021. \textit{Left:} The tgVVC prototype has five thin substrates in sequence and has anti-reflection coating on the outside. The amplitude mask is glued to the forked-grating pattern to cover the central defect. The thee substrates with a liquid-crystal film (substrate 2,3, and 4) are ordered keeping the 2:1 distance ratio between the gratings in mind. \textit{Right:} Images of the two prototypes labeled 1 and 2, in addition to images between crossed polarizers. }
   \end{figure} 
We minimize the grating pitch within the direct-write constraints to maximize the diffraction angle of polariziation leakage, reducing the impact of these diffracted polarization leakage (apodized) pupils.
With the selected grating pitch of 12.65 $\mu$m for the first geometric phase hologram (GPH1) and the IACT F-number, the pupils will be separated by at least 1.5 pupil distance at the Lyot stop.
The other grating periods follow the 1:3:2 ratio of the triple-grating concept and their periods are presented in Table \ref{tab:proto}.
All GPH patterns are generated with 1 micron pixels and we use a charge 6 vortex to decrease the sensitivity to low-order aberrations \cite{ruane2018vortex}. 
    \begin{table}[ht]
\caption{\label{tab:proto} Properties of the five substrates that make the compound triple-grating vector-vortex coronagraph}
\begin{tabular}{c|c|c|c|c|c|c} 
Substrate & Material & Diameter & Thickness & AR-coating & Description & Specification \\
\hline
1 & UVFS & 20 mm & 1 mm & BBAR-A & Amplitude mask & 10 $\mu$m dot \\
2 & UVFS  & 25.4 mm & 1 mm & - & GPH1 & VVC6$+$PG,  12.65 $\mu$m pitch \\
 3 & UVFS  & 25.4 mm & 1 mm & - & GPH2 & PG, 4.217 $\mu$m pitch \\
4 & UVFS & 25.4 mm & 1 mm & - & GPH3 & PG, 6.325 $\mu$m pitch \\
5 & UVFS  & 25.4 mm & 1 mm & BBAR-A & End cap & - \\
\end{tabular}
\end{table} 

We select a single-layer liquid-crystal recipe with a half-wave retardance at 633 nm that are deposited on Thorlabs 1 inch WG41010 substrates.
The triple-gratings consist of five of these substrates, one substrate with an amplitude dot to cover the central defect, three substrates with the patterned liquid-crystal films, and an end-cap to cover the last liquid-crystal film.
Both outer surfaces are coated with anti-reflection coating.
The full sandwich is shown in Figure \ref{fig:f3} and is 5.5 mm thick.
Undersizing the first substrate with the amplitude dot was necessary to accurately align the 10 $\mu$m amplitude dot with the central defect of the first liquid-crystal film. 
Around four mm was removed from the edges of the liquid-crystal films to enable glass-to-glass bonding of the Norland NOA-61 adhesive. 
All three polarization gratings are carefully aligned in clocking to overlap the main diffraction beams before curing the adhesive. 
The NOA-61 epoxy was aged for ~12 hours at 50 degrees Celsius. \\
 \begin{figure} [ht]
   \begin{center}
     \includegraphics[width = \linewidth]{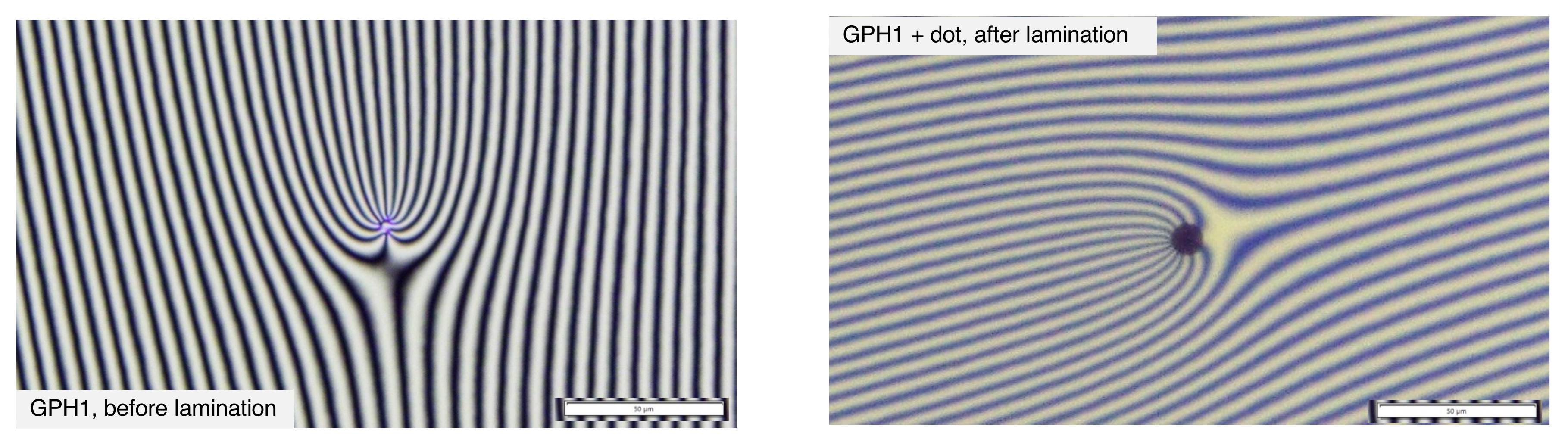}
   \end{center}
   \caption[example] 
%>>>> use \label inside caption to get Fig. number with \ref{}
   { \label{fig:f4} Microscope images of the central part of the first substrate (GPH1) with the forked-grating pattern between crossed polarizers. \textit{Left:} We see exquisite patterning without any defects and a central defect smaller than 5 $\mu$m. \textit{Right:} The same area after being covered by a chromium dot with a 10 $\mu$m diameter.  }
   \end{figure} 
Images of the tgVVC prototypes are shown in Figure \ref{fig:f3}, in addition to images of the optics between crossed polarizers. 
The crossed polarizer images should generate a vortex charge 6 pattern without any residual polarization grating pattern in the vertical direction. 
We conclude that there is a slight rotational alignement error between the gratings, resulting in this pattern. 
In Section \ref{sec:discussion} we will go into manufacturing errors in more detail. 
Additional quality checks of the two tgVVC prototypes have been performed by ImagineOptix. 
Zero-order leakage measurements as function of wavelength show that the retardance should be close to half-wave at 633 nm, and far-field diffraction errors indicate the presence of additional non-standard diffraction orders.
All main diffraction orders agree with the grating pitch, however, the non-standard diffraction orders include half-orders ($\pm 0.5,\pm 1.5$), with a separation that is between the main orders. 
Overall, we find multiple clues of the presence of manufacturing errors that could impact the performance of the optic. \\
Therefore, we explore the coronagraphic performance with these possible errors in mind, while also separately look more into the manufacturing errors. 
ImagineOptix has produced a third optic called the 'bonus part' consisting of only GPH1 and an amplitude mask, that is ideal for studying the manufacturing errors. 
Crossed polarizer microscope images of the amplitude dot alignment of this bonus part are shown in Figure \ref{fig:f4}.
The images show no patterning defects except for the central defect, which is extremely small with a diameter of roughly $5\mu$m. 
The patterning quality looks good, although this is hard to verify with these images, and the alignment of the central dot is close to perfect. 

\section{Performance of the tgVVC prototypes}
\label{sec:contrast}

We test both tgVVC prototypes from ImagineOptix at two different in-air coronagraphic testbeds.
The testbeds are the In-Air Coronagraphic Testbed\cite{baxter2021design} (IACT) at NASA's JPL and the Space Coronagraphic Optical Bench (SCoOB) at the University of Arizona Space Astrophysics Lab\cite{ashcraft2022space,van2022space} (UASAL).
Both testbeds have clear apertures and are optimized for characterizing the performance with VVC coronagraphs. 
Other similarities are the use of a single deformable mirror, the ability to filter polarization leakage, and a limited number of optical elements to improve the performance and keep simplicity. 
As a result, the wavefront control only allows for the digging of one-sided dark zones, although this is not an issue for the prototype testing.
The differences are in the wavefront control, the use of stops in the optical system and the F-numbers. 
The IACT is F/82.5 at the focal plane mask, has model-based electric field conjugation\cite{give2007electric,groff2016methods} (EFC) and some field stops reduce the impact of high-order aberrations. The SCoOB testbed is is F/48 at the focal plane mask, uses the data-driven implicit EFC\cite{haffert2023implicit} (iEFC) and did not have stops when the measurements were taken.
The many similarities allow us to compare results, while the differences between the two testbeds enable us to differentiate between testbed effects or limitations of the tgVVC.\\
\begin{figure} [ht]
   \begin{center}
     \includegraphics[width = \linewidth]{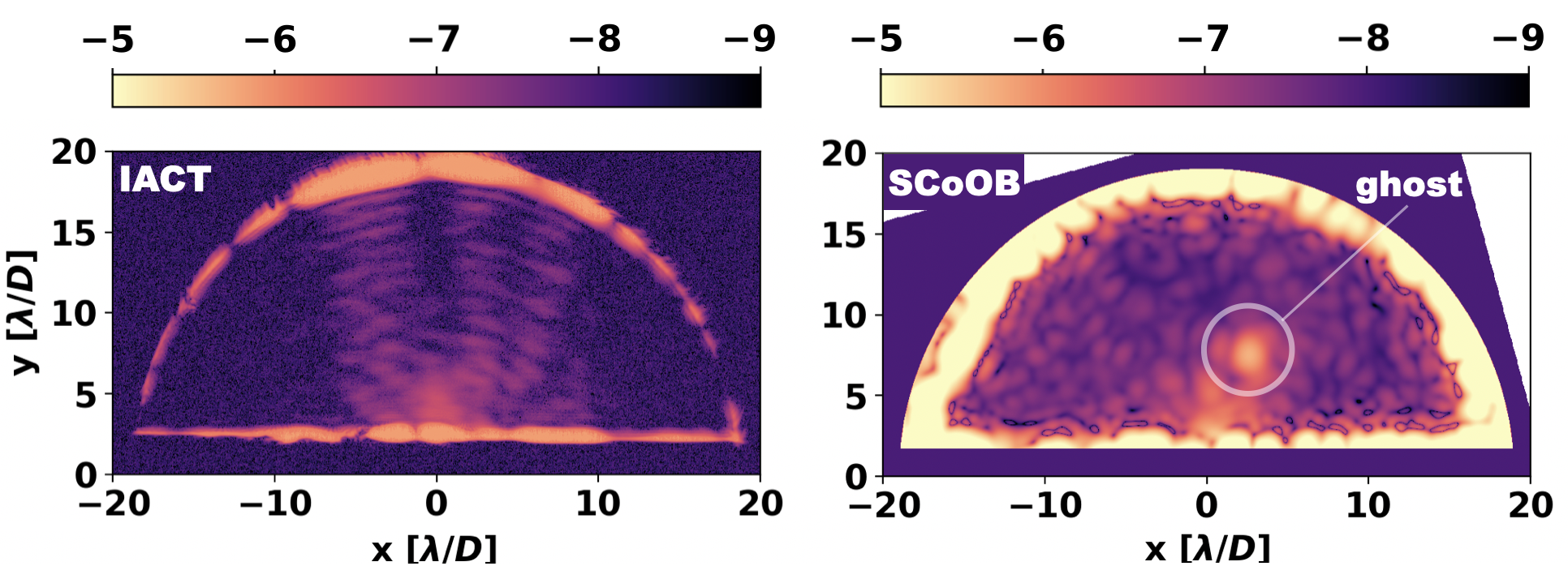}
   \end{center}
   \caption[example] 
%>>>> use \label inside caption to get Fig. number with \ref{}
   { \label{fig:f5} Dark hole measurements of the tgVVC1 prototype measured with monochromatic light at 633 nm and polarization filtering in two high-contrast imaging testbeds. \textit{Left:}. The dark hole image of the tgVVC1 prototype as measured in the In-Air Coronagraphic Testbed at NASA's Jet Propulsion Laboratory. The D-shaped dark hole has a size of 3-18 $\lambda/D$, light outside this dark hole is blocked by a field stop. Model-based EFC was used to reconstruct the wavefront correction. The average dark zone contrast from 3-14 $\lambda/D$ is $3\times10^{-8}$. \textit{Right:}  The dark hole image of the tgVVC1 prototype as measured in the Space Coronagraphic Optical Bench at the University of Arizona Space Astrophysics Lab. The D-shaped dark hole has a size of 3-14 $\lambda/D$ and no field stop is present. Implicit EFC was used to reconstruct the wavefront correction. The bright speckle is a ghost from the quarter-wave plate and the average dark zone contrast without this ghost is 6$\times10^{-8}$. }
   \end{figure} 
Measurements of the second prototype, tgVVC2, on the IACT testbed revealed that the performance is worse by a factor of a few.
Therefore, no tests of the tgVVC2 prototype were conducted on the SCoOB testbed. 
A likely cause for the reduced performance is an unexpected diffracted beam at 45 degrees from the diffraction direction with a separation less than a quarter of the first grating order.
We focus on the tgVVC1 prototype and the results at the IACT and the SCoOB testbed is shown in Figure \ref{fig:f5}. 
On the IACT the dark hole is 3-18 $\lambda/D$ and has a mean contrast of $2.1\times10^{-8}$ in this full dark zone. 
The EFC was combined with the beta-bumping technique \cite{seo2017hybrid} to reach this performance, with strong regularization as described in Ruane et al. 2022 \cite{ruane2022broadband} 
There is a lot of structure in this dark hole, with a ringing structure in three main stripes extending in the vertical direction, while the corners of the D-shaped dark hole contain almost no speckles.
The speckle structure in the dark zone on the SCoOB testbed is completely different. 
The D-shaped dark hole extends from 3-14 $\lambda/D$, contains a ghost from the quarter-wave plate, and has a mostly constant speckle field over the full dark zone.
The SCoOB mean contrast for the tgVVC1 prototype is worse, with an average of 6$\times10^{-8}$ after masking the ghost. \\
Comparing the contrast for similarly shaped dark zones of 3-14 $\lambda/D$, gives a contrast of $3\times10^{-8}$ for IACT compared to the 6$\times10^{-8}$ of the SCoOB testbed, a difference of a factor of two.
With the minor contrast differences and extreme differences in the speckle field, a direct comparison has interesting implications.
The vertical structures in the IACT dark zone are brighter than the speckles of the SCoOB dark zone at the same location.
Therefore, it would seem that these structures are not incoherent structures generated by the tgVVC1 optic, but rather not controlled as consequence of model inaccuracies of the EFC algorithm.
As the EFC model is based on ideal vortex coronagraphic patterns, some differences from manufacturing or patterning inaccuracies might make it difficult to control these speckles with EFC.
Because iEFC is empirically calibrated, it is robust against these situations where optical models might be difficult to realize.
To summarize, the structured IACT speckle field suggests that the tgVVC1 has some manufacturing errors, although these do not have to impact the contrast performance.\\

\section{Discussion}
\label{sec:discussion}

We present the manufactured a triple-grating vector vortex coronagraph prototype and its performance in two high-contrast imaging testbeds. 
While the contrast does reach a contrast level that is expected for a first prototype on in-air testbeds, we find that there are clear signs of manufacturing errors.
First, the crossed polarizer images show residual grating patterns from imperfect grating clocking alignment.
Second, polarization leakage filtering was necessary to reach the reported contrast levels. 
This could point to either pattern writing errors resulting in an increased zero-order leakage or deviations from half-wave retardance.
Third, the dark hole structures in the IACT are a likely result of a model mismatch with a standard vortex pattern, also a sign of pattern writing errors.
Lastly, the far-field diffraction reported by ImagineOptix shows the presence of half-order diffraction. 
These half-orders are also generated when the grating pattern has writing errors. 
Overall, these errors warrant a closer inspection if we are to make improved tgVVC prototypes.
Further study of the diffraction efficiency, retardance, and patterning accuracy with far-field diffraction measurements and imaging Mueller matrix ellipsometry of the Bonus part is required.
   
\section{Conclusion}
\label{sec:conclusion}
\begin{itemize}
\item The multi-grating concept can suppress polarization leakage terms by many orders of magnitude by combining phase patterns with polarization gratings.
\item We apply this concept to the vector vortex coronagraph and successfully manufacture two triple-grating prototypes.
\item We test these triple-grating vector vortex coronagraphs in two high-contrast imaging testbeds and demonstrate good performance.
\item On the In-Air Coronagraphic Testbed at NASA's Jet Propulsion Laboratory we report a contrast of $2.1\times10^{-8}$ in a D-shaped dark hole of 3-18 $\lambda/D$.
\item On the Space Coronagraphic Optical Bench at the University of Arizona Space Astrophysics Lab we report a contrast of $6\times10^{-8}$ in a D-shaped dark hole of 3-14 $\lambda/D$.
\item A comparison between these two dark holes suggests that the IACT performance was limited by EFC model errors. 
\item We find that there are clear signs of manufacturing errors of the prototype that can easily be mitigated by more controlled manufacturing. We will do this in collaboration with ColorLink Japan, ltd. 
\end{itemize}

\acknowledgments % equivalent to \section*{ACKNOWLEDGMENTS}       
We are very grateful for support from the Netherlands Space Office. 

% References
\bibliography{TGVVC} % bibliography data in report.bib
\bibliographystyle{spiebib} % makes bibtex use spiebib.bst

\end{document}